# Incompatible Liquids in Confined Conditions

Krisztina László[1*], Bruno Demé[2], Orsolya Czakkel[2], Erik Geissler[3,4]

[1]Department of Physical Chemistry and Materials Science, Budapest University of Technology and Economics, H-1521 Budapest, Hungary

[2]Institut Laue Langevin, CS 20156, 71 rue des Martyrs, 38042 Grenoble Cedex 9 France

[3]University Grenoble Alpes, LIPhy, F-38000 Grenoble, France

[4]CNRS, LIPhy, F-38000 Grenoble, France

**ABSTRACT**

In applications involving organic vapour, the performance of high surface area carbons is often challenged by water vapour in the atmosphere. Small angle neutron scattering (SANS), through its ability to distinguish between different components by means of contrast variation, is ideally suited to investigating the behaviour of adsorbed layers in simultaneous contact with a mixed vapour phase. Even at high relative humidity (RH), water alone forms a discontinuous film composed of clusters on the surface of the oxidized microporous carbon used for these studies. When toluene is also present, all the available carbon surface is wetted. Toluene and water adsorb as a single phase already at RH 11.5%, and the concentration of water present in the adsorbed phase is as high as 2.9 wt.%, far above its solubility in bulk toluene (0.033 wt.% at 25 ºC). At RH 87% the concentration of water in the adsorbed phase is four times higher, approximately 12 wt.%. The recently proposed mechanism of anchoring of the water by the aromatic molecules may provide an explanation for this phenomenon.

# INTRODUCTION

The separation of volatile organic vapours from air is still to this day one of the leading applications of high surface area nanostructured carbons. As such carbons are intrinsically general adsorbents, under ambient conditions the relative humidity of air can challenge the performance of the carbon bed. It is therefore of some importance that relevant information on adsorption be sought in experiments other than those where the carbon is exposed just to a single component vapour. Investigations of the adsorption from binary vapour phase are most often performed under dynamic conditions, i.e., in column experiments [1-5] or by successively exposing the carbon bed to the vapours in question [2, 6]. In the latter case it was found that the sequence of the vapours may fundamentally affect the development of the adsorbed layer.

Small angle neutron scattering (SANS), through its ability to distinguish between different components by means of contrast variation, is ideally suited to investigating the behaviour of the adsorbed bed equilibrated in mixed vapour phase [7, 8]. The contrast variation is achieved by employing mixtures of deuterated and protonated solvents in different proportions.

Incompatible liquids, such as polar and aromatic compounds, have very low mutual solubility in the bulk state. Under confined conditions, however, earlier SANS observations on adsorption in high surface area carbons from mixtures of water and toluene [7] detected an appreciable amount of water mixed with the toluene in the adsorbed phase, greatly in excess of their solubility in the bulk state (solubility of toluene in water: 0.052 wt.% and water in toluene: 0.033 wt.%, both at 25 ºC [9]). In the context of microporous systems, it is noteworthy that the pressures generated by surface tension inside liquid droplets of nanometre size are of the order of 100 MPa. Sawamura *et al*. [10] have studied the effect of pressure on the solubility of various alkylbenzene compounds in bulk water up to 400 MPa in the 0 - 50 ºC temperature range, and found that generally solubility increases initially up to a pressure of about 200 MPa, but decreases thereafter. For toluene in water, however, the increase is only modest, rising from about $1.09 \times 10^{-4}$ to $1.28 \times 10^{-4}$ molar fraction at 25 ºC. The explanation of the increased miscibility of water and toluene in confined spaces therefore lies elsewhere. The question is currently of importance since enhanced interactions occurring in confinement under ambient conditions may open new avenues in chemical synthesis [11, 12].

Here we report a structural investigation by SANS combined with contrast variation of adsorption on a microporous carbon bed, in which the pore sizes are of the order of a few nanometres. The high surface area carbon was exposed to an atmosphere containing simultaneously a model aromatic molecule, toluene, and water vapour.

**MATERIALS AND METHODS**

A nanostructured carbon (Norit R1 Extra exposed to concentrated nitric acid for 3 h at room temperature) [13] was used in the experiments (Table 1). The ash content of the carbon is 2.2 wt. % and the surface O/C atomic ratio was found by XPS to be 10.4 %. The powdered samples were inserted into low boron content glass tubes of 5 mm outer diameter and placed in contact simultaneously, on the one hand, with the vapour of different mixtures of $H_2O$ and $D_2O$ and, on the other hand, with the vapour of different mixtures of protonated and deuterated toluene. Two values of relative humidity (RH) were used for the water, 11.3 % (saturated solution of LiCl), and 87 % (saturated solution of KCl) at 20 ºC. The samples were allowed to come to equilibrium under thermostated conditions at 20 ºC for 2 months. The SANS measurements were made on the D16 instrument at the Institut Laue Langevin (ILL), Grenoble, at wavelength 4.75 Å, with wavelength spread $\Delta\lambda/\lambda=0.01$. Corrections for incoherent scattering were made by comparing SANS spectra with those obtained from small angle X-ray scattering (SAXS) measurements, performed at the European Synchrotron Radiation Facility, Grenoble.

The proton/deuteron content of the samples is denoted by the nomenclature CW$n$T$m$, where W stands for water, T for toluene, and $n$ and $m$ take the values 1- 4. Thus $n$=1 is 50% $H_2O$ + 50% $D_2O$, while $n$=4 is pure $D_2O$, with $n$=2 and $n$-3 being the equally spaced intervening compositions. The same nomenclature is used for toluene, with T4 being pure $C_7D_8$. Thus CW4T4 is the carbon sample that has been exposed simultaneously to the vapours of 100% $D_2O$ and 100% toluene-D [7].

**Table 1** Characteristic data of the high surface area carbon*

| | |
|---|---|
| $S_{BET}$, m$^2$/g | 1450 |
| $V_{TOT}$, cm$^3$/g | 0.70 |
| $V_{0.95,N2}$ cm$^3$/g | 0.67 |
| $W_{0,N2}$ cm$^3$/g | 0.58 |
| $E$, kJ/mol | 15.0 |
| $w$, Å | 17.4 |
| $d_{av}$, Å | 19.4 |
| $\delta_{He}$, g/cm$^3$ | 2.10 |
| $\delta_{Xylene}$, g/cm$^3$ | 2.10 |
| $\delta_{av}$, g/cm$^3$ | 0.835 |
| $S_X$, m$^2$/g | 1690 |

*$S_{BET}$: surface area obtained by applying BET model to the nitrogen adsorption data, $V_{TOT}$: total pore volume at $p/p_0 \approx 1$, $V_{0.95,N2}$ total pore volume at $p/p_0 \approx 0.95$, $W_{0,N2}$: micropore volume from DR plot, $E$: characteristic energy of adsorption, $w$: average width of the micropores, $d_{av}$: average pore diameter, $\delta_{He}$: helium density of the carbon matrix, $\delta_{Xylene}$: density of carbon matrix measured with xylene, $\delta_{av}$: apparent density, $S_X$: specific surface area from SAXS measurements. Note that for consistency with the X-ray results the pore sizes are expressed in ångström units.

Water vapour adsorption isotherms of this carbon were measured using a volumetric Hydrosorb apparatus (Quantachrome) at 20 °C, with vapour generated at 100 °C (**Figure 1**). The sample was previously degassed for 24 hours at 20 mTorr at room temperature, and the adsorption/desorption curves were completed within 13 hours.

**RESULTS AND DISCUSSION**

The volumetric water adsorption isotherm shows that at 20º C the carbon adsorbs 6.3 mg/g and 241 mg/g water at RH 11.3 and 87 %, respectively. However, as has been previously observed by wide angle X-ray scattering (WAXS) measurements, the water uptake is substantially greater when the microporous carbon is exposed to water vapour for extended periods, reaching 27.2 mg/g and 642 mg/g, respectively [13] (**Figure 1** open circles).

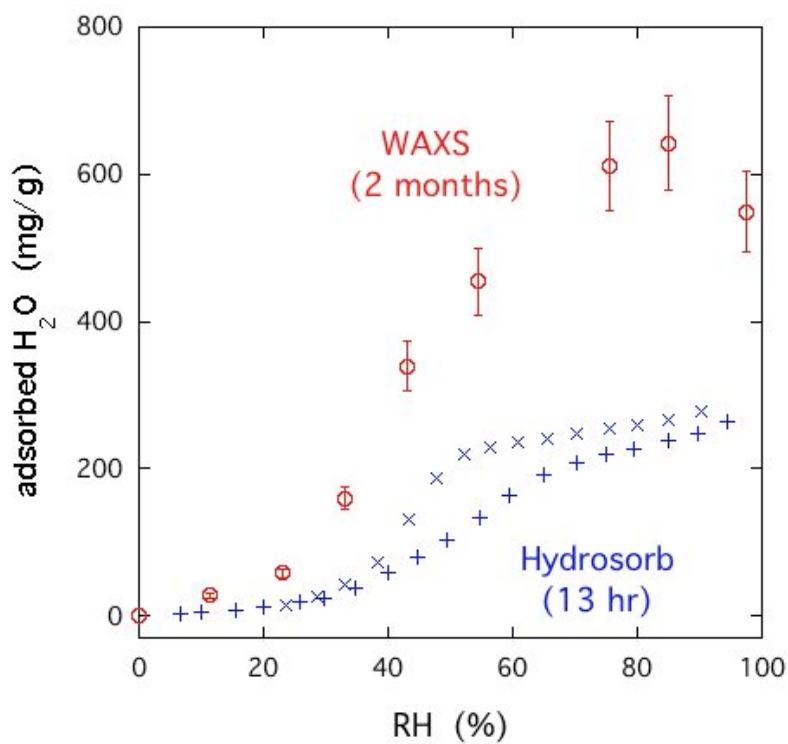

**Figure 1**. Water uptake by the carbon sample in a 13-hour period from Hydrosorb measurements, adsorption (+), desorption (×), compared with that measured by WAXS after exposure for 2 months (o).

In the SANS measurements, the total detected intensity $I_{tot}(q)$ is the sum of the coherent signal $I(q)$ characteristic of the system investigated and a constant incoherent intensity $I_{inc}$. The momentum transfer $q$ is equal to $(4\pi/\lambda)\cdot\sin(\theta/2)$, where $\lambda$ is the wavelength of the incident neutrons and $\theta$ is the scattering angle. $I(q)$ is defined by the relative positions of the atoms in the sample and their contrast factor $(\rho_j-\rho_k)^2$, where $\rho_j$ is the scattering length density of the nuclear species in question and $\rho_k$ that of its surroundings. The incoherent intensity $I_{inc}$ is essentially proportional to the number of protons in the sample. Filling the pores with an adsorbate changes the contrast. **Figure 2** illustrates the reduction in the coherent SANS signal intensity $I(q)$ of the carbon due to contrast matching by the adsorbate when either pure deuterated toluene $C_7D_8$ vapour ($p/p_0$=0.028) or pure $D_2O$ (RH=87%) is adsorbed [14]. This intensity reduction, which is also related to the penetration of the vapour molecules into the pores, occurs for $q$ less than about 1 Å$^{-1}$, where the material appears to be continuous to the neutrons. On the contrary, in the wide angle scattering region ($q > 1$ Å$^{-1}$), the discrete atomic nature of the medium is probed, and the signal of the adsorbed molecules adds to that of the carbon. **Figure 2** also shows that with $D_2O$ in the lowest $q$ range the signal intensity increases, with $I(q) \propto 1/q^m$, where $m>3$. This is a tell-tale sign of surface scattering: wetting of the internal surfaces of the pores by the water molecules is incomplete, and the contrast factor between the carbon and gas phase, $(\rho_C-\rho_{gas})^2$, is reduced to $(\rho_C-\rho_{D2O})^2$ only at limited sites. A similar effect is found in other carbons when the water does not wet the surface, and adsorption on the hydrophobic surface proceeds by nucleation and clustering [15]. By contrast, with toluene the power law slope in the low $q$ region is characteristic of volume scattering. When both water and toluene are adsorbed by the carbon, a ternary system results. The coherent scattering intensity can then be expressed in terms of the partial structure factors $S_{jk}(q)$, where the subscripts $j$ and $k$ are $C$ (carbon), $w$ (water), or $tol$ (toluene), thus

$$I(q)=(\rho_C-\rho_{tol})^2 S_{CC}(q) + (\rho_{tol}-\rho_w)^2 S_{ww}(q) + (\rho_C-\rho_{tol})(\rho_C-\rho_w) S_{Cw}(q) \qquad (1)$$

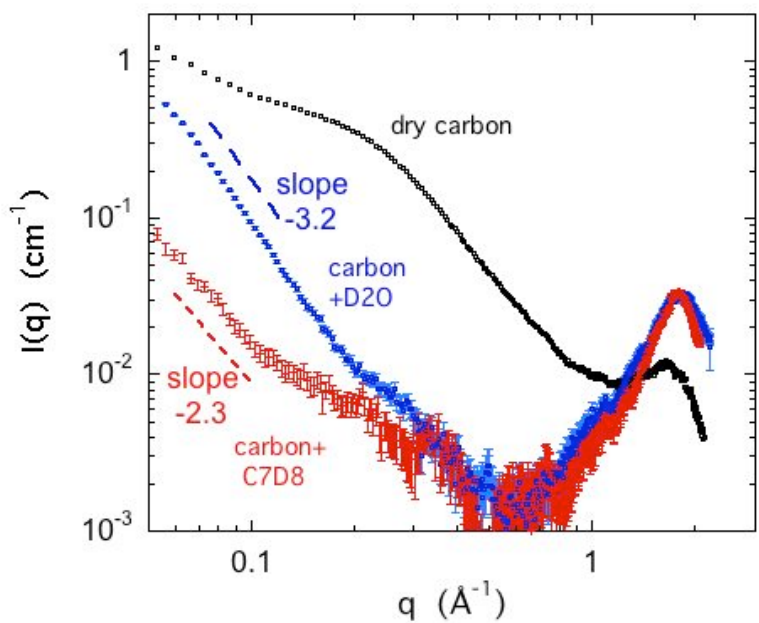

**Figure 2.** Coherent SANS response of the dry carbon (black), carbon with $D_2O$ vapour at 87% RH (blue), and carbon with toluene-D vapour in equilibrium with the bulk liquid at 20ºC (red). Dashed lines indicate the power law slopes at low $q$ (-3.2 and -2.3, respectively).

In Eq. 1, the $\rho_j$ are the neutron scattering length densities of the corresponding components, in which

$$\rho_j=(N_A \Sigma b_j/M_j)d_j \qquad (2)$$

where $N_A$ is Avogardo's number, $\Sigma b_j$ the sum of the neutron scattering lengths of the nuclei in the component, $M_j$ its molar mass and $d_j$ is its mass density. By varying the proportion of $H_2O$ and $D_2O$ in the water, and of $C_7H_8$ and $C_7D_8$ in the toluene, Eq. 1 yields a set of different SANS responses $I(q)$ that may be solved for the partial structure factors $S_{jk}(q)$. Strictly speaking, Eq. 1 should include further coherent scattering terms describing correlations between deuterated and hydrogenated components of the solvent. These terms are defined by the mean square amplitude of the fluctuations, $<\Delta\varphi^2>=\varphi(1-\varphi)$, where $\varphi$ is the volume fraction of either one of the components. For bulk water in the SANS region $q<1$ Å$^{-1}$, for example, the corresponding term can be expressed in the form

$$I_{H2O\,D2O}(q)=(\rho_{H2O}-\rho_{D2O})^2 v_0\, \varphi(1-\varphi)/N_A \qquad (3)$$

in which $v_0$ is the molar volume of water and φ is the volume fraction of $H_2O$ in the mixture. Since the coherent scattering intensities corresponding to Eq. 3 are small, however, and much smaller than the incoherent signal from the protons of the hydrogenated molecule, such contributions are neglected here. Note that the third (cross) term in Eq. 1 is sometimes written with a factor of 2; this definition merely modifies the numerical value of $S_{Cw}(q)$, not its functional behaviour.

In the set of equations represented by Eq. 1, $S_{CC}(q)$ was measured directly from the dry carbon. Physically meaningful positive values of the direct term $S_{ww}(q)$ could be obtained only by assuming that the mass density of the adsorbed phase is less than that in the bulk system, about 95%. This conclusion is corroborated by WAXS observations of water adsorbed on carbon surfaces, which indicate that its structure is liquid-like, but appreciably different from that of bulk water [16]. In the present sample, which contains surface oxygen, the adsorbed water is even more ordered, with well resolved WAXS peaks at 1.85, 3.05 and 5.2 Å$^{-1}$ [13].) It should be noted that at small $q$, where the largest pores are incompletely filled, the system is no longer ternary, but quaternary, and Eq. 1 is hence no longer applicable. At high $q$ the continuous medium assumption loses its validity. For these reasons the region of confidence in the results, shown in **Figure 3**, lies in a limited $q$ range, indicated by the vertical dashed lines in the figure.

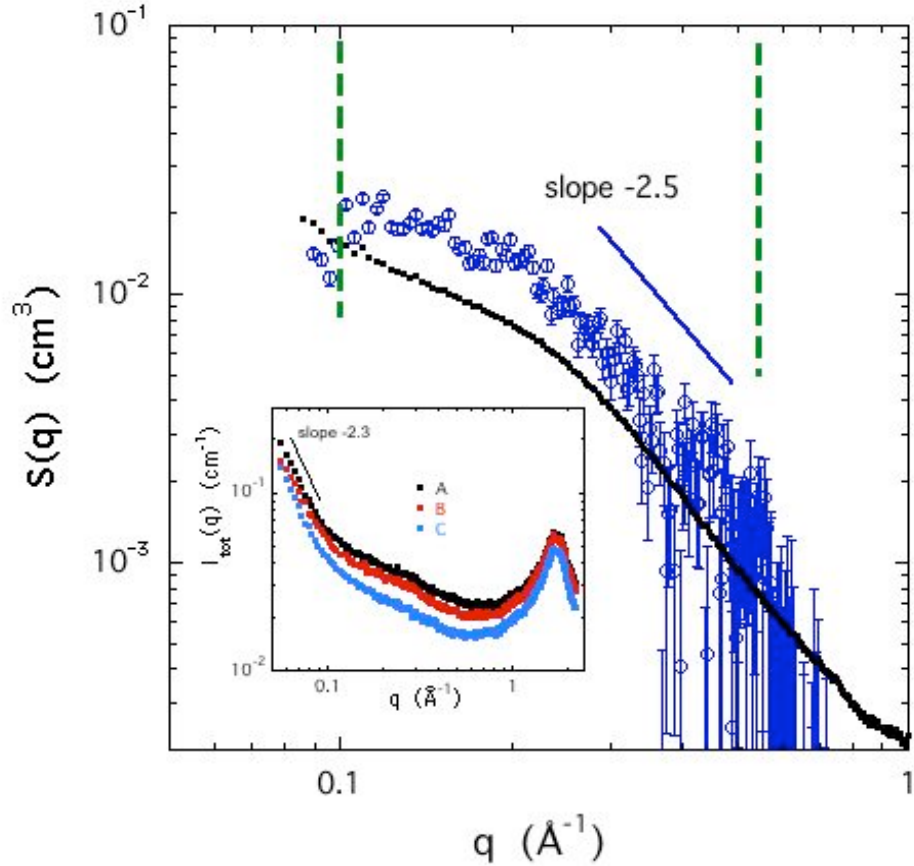

**Figure 3**. Partial structure factors $S_{ww}(q)$ (o) and $S_{cc}(q)$ (•) of the carbon-toluene-water system with RH=87%, as calculated from Eq. 1. The vertical dashed lines delimit the region of confidence in the approximations used for the calculations (from ref. 7). Inset: total SANS intensity $I_{tot}(q)$ in carbon containing: (A) $C_7D_8$ + (0.5$H_2O$+0.5$D_2O$) vapour (CW1T4) at RH 87%; (B) same as A, but with RH 11.5%; (C) $C_7D_8$ + $D_2O$ vapour (CW4T4) at RH 87%. The slope in the power law region at low $q$ is $-(2.3\pm0.1)$.

As noted previously [7], the slope of the partial structure factors in the double logarithmic representation of **Figure 3**, -2.5, is appreciably lower than that of surface scattering. This indicates that the mixed toluene and water adsorbate at conditions of high RH (87%) forms a single phase, rather than two separate phases with a well defined interface. The water content in the toluene can be estimated from the incoherent scattering intensity $I_{inc}$ at different $H_2O/D_2O$ ratios. The value found for the weight fraction, $w=12\pm1\%$, is in excellent agreement with that obtained from the coherent signals $I(q)$ of the same samples.

In the inset of **Figure 3** the total scattering signal $I_{tot}(q)$ from carbon samples CW1T4 exposed to a mixture of fully deuterated toluene-D vapour and water vapour of composition 50%$H_2O$:50%$D_2O$ is compared at two different values of RH, 87% and 11.3%. The difference between these signals is due mainly to incoherent scattering, i.e., to the protons present in the adsorbed water. For comparison, this inset also shows the signal from the carbon sample CW4T4, with $D_2O$ and toluene-D vapour. In each of these samples the slope of the power law region at low $q$ is -2.3, i.e., characteristic of volume scattering. The disappearance of the surface scattering feature in this region implies that the toluene molecules cover the sites on the carbon surface that the water molecules alone failed to wet. In **Figure 4** the difference in the incoherent intensity $\Delta I_{inc}$ from that in the corresponding fully deuterated sample, CW4T4, is plotted as a function of (protonated) $H_2O$ content in the water bath. The slope of the linear plot through the data points is roughly four times smaller with RH 11.3% than for RH 87%. It follows that the water content in the mixed adsorbate with RH 11.3% is about one quarter of that with RH 87%, i.e., 2.9 wt.%. However, even the latter value is still two orders of magnitude greater than the solubility of water in bulk toluene, i.e., 0.033 wt.% at 25 ºC [9].

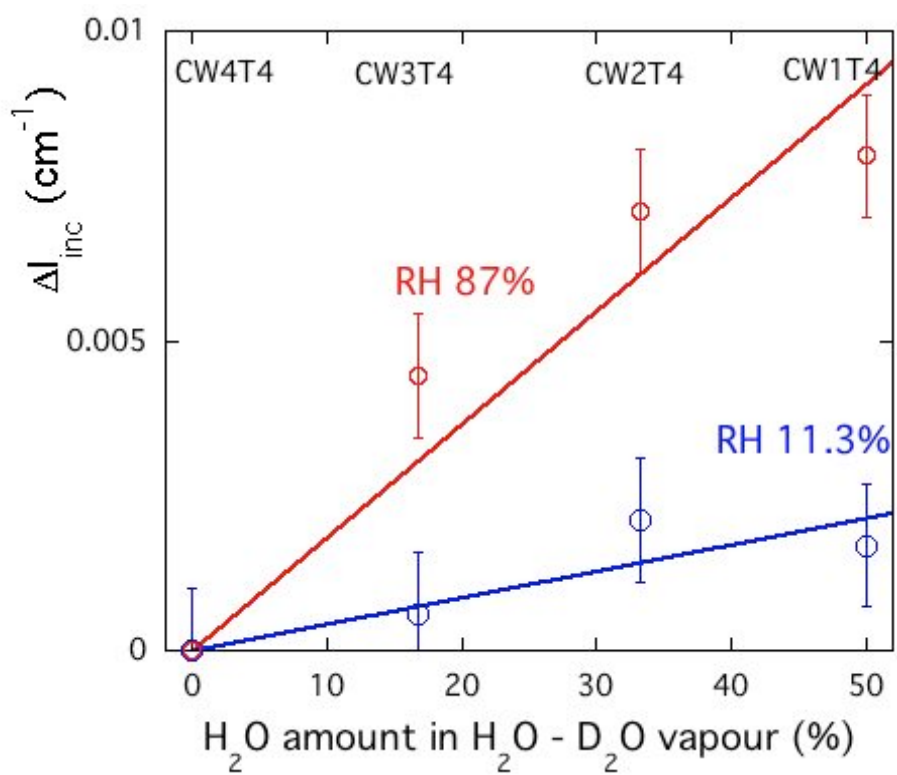

**Figure 4.** Variation of incoherent intensity scattered by samples with adsorbed toluene and water vapour, as a function of the $H_2O$ percentage in the water vapour ($H_2O + D_2O$).

It is instructive to compare the effect of adsorption of water in the presence of toluene on the scattering response as discussed above with that of water vapour alone. **Figure 5** shows the signals from pure carbon and that from CW4 at RH 87%. In the SANS region $q<1$ Å$^{-1}$ the response of the latter is strongly reduced with respect to the dry carbon, owing to the decrease in contrast factor from $\rho_C^2$ to $(\rho_C-\rho_{D2O})^2$. The upper horizontal scale in this figure is a spatial length scale defined by $2.5/q$. (The numerical factor $2.5 \approx \sqrt{2\pi}$ in this relationship is a convenient approximation for designating particle or pore sizes, as opposed to $2\pi/q$ which defines distances of separation.) The behaviour at RH 87% in **Figure 5** differs markedly from that at RH 11.3%, where, in the lower $q$ region, the signal from the adsorbate-containing samples lies *above* that of the dry carbon. When the micropores in the carbon are only partially filled, at long distance scales (i.e., at low $q$), the apparent mass density (which, according to Eq. 2, is proportional to the scattering length density) of the carbon is far smaller than that at short distances. As the density of the sample is defined by the size of the probe, at short distances the effective density, measured by helium gas, is the helium density, i.e., the density of the pores walls. At low $q$ the adsorbed $D_2O$ enhances the apparent density of the carbon and thus increases the contrast with the air in the pores, thereby increasing the signal. At short length scales (high $q$ values), on the other hand, the effective contrast is between the pore filling water and the carbon, and the signal is accordingly reduced. For the RH 11.3% sample, therefore, the main effect of water adsorption becomes visible in the highest $q$ range, just where the continuous medium assumption breaks down, and accordingly it is difficult to estimate the degree of filling for this system. An estimate of the total amount of water can however be obtained from the wide angle scattering peak of water just below 2 $\text{Å}^{-1}$. Note that the scattering response of the adsorbed water differs appreciably from that of the bulk liquid [16, 17]: the principal peak is located at about 1.85 $\text{Å}^{-1}$, while in bulk liquid water it is close to 2.0 $\text{Å}^{-1}$. From the relative intensities of these peaks it can be estimated that the water content in the RH 11.3% sample without toluene is about 16% of that at RH 87%. Although this degree of water adsorption is already appreciable at such a low relative humidity, it is interesting to note that the ratio increases to 25% when toluene is present.

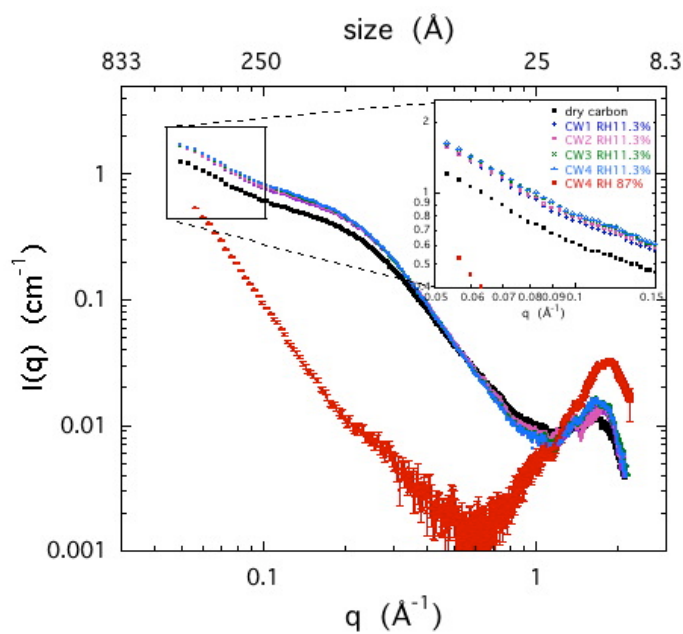

**Figure 5**. Comparison of SANS response of dry carbon (black symbols), carbon with D$_2$O at RH 87% (red symbols with error bars), and carbon with RH 11.3%. at different deuteration ratios. Symbols for RH 11.3% in zoom inset: CW1 (blue +), CW2 (mauve o), CW3 (green ×), CW4 (cyan Δ).

The above findings appear even more striking when expressed in molar terms: at RH 87% the adsorbed water molecules account for 40% of the total number of adsorbed molecules, while at RH 11.3% they constitute 13%. Recent Monte Carlo simulations [18] have shown that in the graphite – benzene – water system the benzene molecules act, somewhat counter-intuitively, as anchors for the water molecules to adsorb and form clusters, thereby enhancing their uptake. In the present system it may therefore be reasonable to expect that the more polar toluene molecules will facilitate this adsorption even more strongly. Such an effect is additional to the standard mechanism of adsorption by formation of water clusters around polar groups on the carbon surface.

**CONCLUSIONS**

SANS combined with contrast variation is a powerful means of investigating the behaviour of otherwise immiscible toluene – water mixtures in confined conditions. Unlike what happens with pure toluene vapour, and in spite of its high surface oxygen content, when the carbon is exposed to pure water vapour, surface wetting by the adsorbate is incomplete. The existence of water-free surfaces even at RH=100% provides direct evidence that the film of adsorbed water is discontinuous and that, by corollary, the water molecules form clusters on the carbon surface and the density of the adsorbed water is lower than that in the bulk liquid. The SANS observations show that when toluene and water vapours are simultaneously present, all the available carbon surfaces are completely wetted. Already at RH 11.3%, both molecules adsorb on this oxidized high surface area carbon as a single phase. Even at this low RH, the fraction of water present is as high as 2.9 wt.%, far above its solubility in bulk toluene (0.033 wt.% at 25 ºC). At RH 87% the overwhelming majority of the mass of adsorbed phase consists of toluene (the concentration of water in this phase is approximately 12 wt.%), but in molar terms the number of adsorbed water molecules account for 40% of the total. The mechanism of anchoring of the water by the aromatic molecules, recently proposed by Do et al. [18], may provide an explanation for this phenomenon.

**ACKNOWLEDGEMENT**

We extend our warm thanks to D.D. Do for enlightening discussions. The ILL is gratefully acknowledged for access to the instrument D16, and the ESRF for access to BM02. The research was supported by OTKA K109558.

**Table of Contents (TOC) Image**

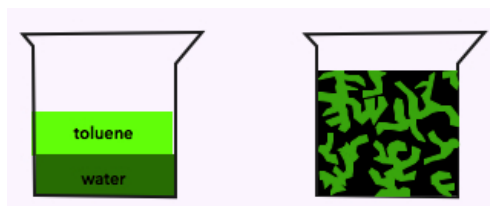